\documentclass[12pt]{article}
\usepackage{axodraw}
\newcommand{\mysection}{\setcounter{equation}{0}\section}

\def\beq{\begin{equation}}
\def\eeq{\end{equation}}
\def\beqa{\begin{eqnarray}}
\def\eeqa{\end{eqnarray}}
 
\newlength{\dinwidth} \newlength{\dinmargin}
\begin{document}
 
\begin{center}
{\Large \bf 
Status of Eikonal Two-Loop Calculations\\ with Massive Quarks}
\end{center}
\vspace{2mm}
\begin{center}
{\large Nikolaos Kidonakis\footnote{Presented at DIS 2008, London, England, 
April 7-11, 2008} 
and Philip Stephens}\\
\vspace{2mm}
{\it Kennesaw State University, Physics \#1202\\
1000 Chastain Rd., Kennesaw, GA 30144-5591}
\end{center}

\begin{abstract}
We present results for two-loop 
diagrams with massive quarks in the eikonal 
approximation. Explicit expressions are given for the UV poles 
in dimensional regularization of several of the required 
integrals.
\end{abstract}

\thispagestyle{empty}  \setcounter{page}{1}

\mysection{Introduction}

The calculation of threshold corrections to hard scattering cross 
sections beyond leading logarithms requires the 
calculation of loop diagrams in the eikonal approximation \cite{url}.
One-loop calculations have been performed for all 
2 $\rightarrow$ 2 partonic processes in heavy quark \cite{KS} and 
jet \cite{KOS} production. 
The soft anomalous dimension matrix $\Gamma_S$ at one-loop 
allows the resummation of soft-gluon corrections at next-to-leading
logarithm (NLL) accuracy \cite{KS}. The exponentiation follows from the renormalization 
group evolution of $\Gamma_S$ and involves the calculation of the 
ultraviolet (UV) poles 
in dimensional regularization of one-loop diagrams with eikonal lines.
To extend resummation to next-to-next-to-leading 
logarithms (NNLL) two-loop calculations are required. For massless 
quark-antiquark scattering the two-loop $\Gamma_S$ was completed in \cite{ADS}.
For heavy quark production, however, the result is not known. In this 
contribution we present several results for two-loop diagrams involved 
in the calculation of the two-loop $\Gamma_S$ for massive quarks.
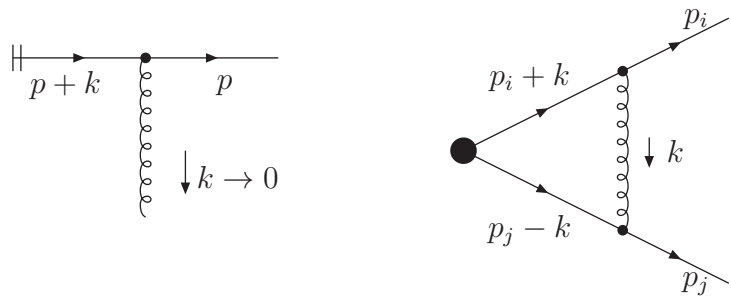
\begin{figure}[htb]
\begin{center}
\begin{picture}(120,120)(0,0)
\Line(0,80)(0,90)
\Line(3,80)(3,90)
\ArrowLine(0,85)(50,85)
\ArrowLine(50,85)(100,85)
\Vertex(50,85){2}
\Gluon(50,25)(50,85){2}{8}
\Text(20,75)[c]{$p+k$}
\Text(80,75)[c]{$p$}
\LongArrow(65,50)(65,35)
\Text(85,40)[c]{$ k \rightarrow 0$}
\end{picture}
\hspace{15mm}
\begin{picture}(120,120)(0,0)
\Vertex(0,50){5}
\ArrowLine(0,50)(60,80)
\ArrowLine(60,80)(100,100)
\Vertex(60,80){2}
\Gluon(60,80)(60,20){2}{8}
\Text(25,78)[c]{$p_i+k$}
\Text(88,100)[c]{$p_i$}
\LongArrow(70,55)(70,45)
\Text(80,50)[c]{$k$}
\ArrowLine(0,50)(60,20)
\ArrowLine(60,20)(100,0)
\Vertex(60,20){2}
\Text(25,20)[c]{$p_j-k$}
\Text(88,0)[c]{$p_j$}
\end{picture}
\end{center}
\caption{The eikonal approximation (left) and a one-loop diagram (right).}
\end{figure}
In the eikonal approximation the usual Feynman rules are
simplified by letting the gluon momentum approach zero (left diagram in Fig. 1):
\beqa
{\bar u}(p) \, (-i g_s T_F^c) \, \gamma^{\mu}
\frac{i (p\!\!/+k\!\!/+m)}{(p+k)^2
-m^2+i\epsilon} \rightarrow {\bar u}(p)\,  g_s T_F^c \, \gamma^{\mu}
\frac{p\!\!/+m}{2p\cdot k+i\epsilon}
={\bar u}(p)\, g_s T_F^c \,
\frac{v^{\mu}}{v\cdot k+i\epsilon}
\nonumber
\eeqa
with $p \propto v$, and $T_F^c$ the generators of SU(3).

\mysection{One-loop and two-loop diagrams}

We perform our calculation for eikonal massive quarks in Feynman gauge
using dimensional regularization with $n=4-\epsilon$.

We begin with the one-loop diagram in Fig. 1. The momentum integral 
is given by 
\beqa
I_{1l} = g_s^2 \int\frac{d^n k}{(2\pi)^n} \frac{(-i)g^{\mu \nu}}{k^2}
\frac{v_i^{\mu}}{v_i\cdot k} \, \frac{(-v_j^{\nu})}{(-v_j\cdot k)} \,.
\nonumber
\eeqa
Using Feynman parametrization, followed by integration over $k$, and
after several manipulations, we find
\beqa
I_{1l}&=&\frac{\alpha_s}{\pi} \, (-1)^{-1-\epsilon/2} \, 2^{5\epsilon/2} \,
\pi^{\epsilon/2} \, \Gamma\left(1+\frac{\epsilon}{2}\right)
(1+\beta^2) \int_0^1 dx \, x^{-1+\epsilon} (1-x)^{-1-\epsilon}
\nonumber \\ && 
\times  \left\{\int_0^1 dz \left[4z \beta^2 (1-z)+1-\beta^2\right]^{-1}
-\frac{\epsilon}{2} \int_0^1 dz \frac{\ln\left[4z \beta^2 (1-z)
+1-\beta^2\right]}{4z \beta^2 (1-z)+1-\beta^2}
+{\cal O}\left(\epsilon^2\right)\right\} 
\nonumber
\label{I1lc}
\eeqa
where $\beta=\sqrt{1-4m^2/s}$.
The integral over $x$ contains both UV and infrared (IR) singularities. 
We isolate the UV singularities, 
$\int_0^1 dx \, x^{-1+\epsilon} \, (1-x)^{-1-\epsilon}
=\frac{1}{\epsilon}+{\rm IR}$,
and find the UV pole and constant terms at one loop:
\beqa
I_{1l}^{UV}&=&\frac{\alpha_s}{\pi} \frac{(1+\beta^2)}{2\beta}
\left\{\frac{1}{\epsilon} \ln\left(\frac{1-\beta}{1+\beta}\right)
+\frac{1}{2} \left(4 \ln 2+\ln \pi-\gamma_E-i\pi\right)
\ln\left(\frac{1-\beta}{1+\beta}\right) \right.
\nonumber \\ && \hspace{23mm} \left.
{}+\frac{1}{4} \ln^2(1+\beta)-\frac{1}{4} \ln^2(1-\beta)
-\frac{1}{2} {\rm Li}_2\left(\frac{1+\beta}{2}\right)
+\frac{1}{2} {\rm Li}_2\left(\frac{1-\beta}{2}\right) \right\} \, .
\nonumber
\label{I1l}
\eeqa
More details on this one-loop integral are given in \cite{KSS}.
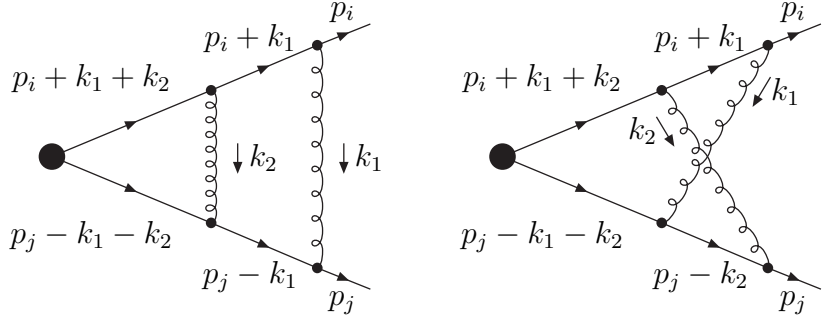
\begin{figure}[htb]
\begin{center}
\begin{picture}(120,120)(0,0)
\Vertex(0,60){5}
\ArrowLine(0,60)(60,85)
\ArrowLine(60,85)(100,102)
\ArrowLine(100,102)(120,110)
\Vertex(60,85){2}
\Vertex(60,35){2}
\Gluon(60,85)(60,35){2}{8}
\Text(15,90)[c]{$p_i+k_1+k_2$}
\Text(75,105)[c]{$p_i+k_1$}
\Text(110,115)[c]{$p_i$}
\LongArrow(70,65)(70,55)
\Text(80,60)[c]{$k_2$}
\ArrowLine(0,60)(60,35)
\ArrowLine(60,35)(100,18)
\ArrowLine(100,18)(120,10)
\Vertex(100,102){2}
\Vertex(100,18){2}
\Gluon(100,102)(100,18){2}{8}
\Text(15,30)[c]{$p_j-k_1-k_2$}
\Text(75,15)[c]{$p_j-k_1$}
\Text(110,5)[c]{$p_j$}
\LongArrow(110,65)(110,55)
\Text(120,60)[c]{$k_1$}
\end{picture}
\hspace{15mm}
\begin{picture}(120,120)(0,0)
\Vertex(0,60){5}
\ArrowLine(0,60)(60,85)
\ArrowLine(60,85)(100,102)
\ArrowLine(100,102)(120,110)
\Vertex(60,85){2}
\Vertex(60,35){2}
\Gluon(60,85)(100,18){2}{8}
\Text(15,90)[c]{$p_i+k_1+k_2$}
\Text(75,105)[c]{$p_i+k_1$}
\Text(110,115)[c]{$p_i$}
\LongArrow(58,75)(64,65)
\Text(53,70)[c]{$k_2$}
\ArrowLine(0,60)(60,35)
\ArrowLine(60,35)(100,18)
\ArrowLine(100,18)(120,10)
\Vertex(100,102){2}
\Vertex(100,18){2}
\Gluon(100,102)(60,35){2}{8}
\Text(15,30)[c]{$p_j-k_1-k_2$}
\Text(75,15)[c]{$p_j-k_2$}
\Text(110,5)[c]{$p_j$}
\LongArrow(101,90)(95,80)
\Text(107,85)[c]{$k_1$}
\end{picture}
\end{center}
\caption{Two loop diagrams with two-gluon exchanges.}
\end{figure}
We now continue with the two-loop diagrams (these are the eikonal versions 
of the diagrams involved in the calculation of the two-loop heavy quark form 
factor \cite{HQFF}). In Fig. 2, we show a diagram 
with two gluons exchanged between the massive quarks (left) and the crossed 
diagram (right). 
We denote by $I_1$ the integral for the first diagram and by $I_2$ that for the 
crossed diagram. We have 
\beqa
I_{1}=g_s^4 \int\frac{d^n k_1}{(2\pi)^n}\frac{d^n k_2}{(2\pi)^n}
\frac{(-i)g^{\mu\nu}}{k_1^2} \frac{(-i)g^{\rho\sigma}}{k_2^2}
\frac{v_i^{\mu}}{v_i\cdot k_1} \frac{v_i^{\rho}}{v_i\cdot (k_1+k_2)}
\frac{(-v_j^{\nu})}{-v_j\cdot k_1} \frac{(-v_j^{\sigma})}{-v_j\cdot
(k_1+k_2)} \, .
\nonumber 
\eeqa
We note that $I_1$ is symmetric under $k_1 \leftrightarrow k_2$ as
is the integral for the crossed diagram, $I_2$. Utilizing the properties 
of these two integrals and the one-loop integral, $I_{1l}$, we find the 
relation
\beqa
I_1=\frac{1}{2} (I_{1l})^2-I_2 \, .
\nonumber
\eeqa
Therefore $I_1$ is determined once we calculate $I_2$.
For the crossed diagram, we have
\beqa
I_{2}=g_s^4 \int\frac{d^n k_1}{(2\pi)^n}\frac{d^n k_2}{(2\pi)^n}
\frac{(-i)g^{\mu\nu}}{k_1^2} \frac{(-i)g^{\rho\sigma}}{k_2^2}
\frac{v_i^{\mu}}{v_i\cdot k_1} \frac{v_i^{\rho}}{v_i\cdot (k_1+k_2)}
\frac{(-v_j^{\nu})}{-v_j\cdot (k_1+k_2)} \frac{(-v_j^{\sigma})}
{-v_j\cdot k_2} \, .
\nonumber
\eeqa
We begin with the $k_2$ integral and after some work find 
\beqa
I_2&=& -i \, \frac{\alpha_s^2}{\pi^2} \, 2^{-4+\epsilon} \, 
\pi^{-2+3\epsilon/2} \, \Gamma\left(1-\frac{\epsilon}{2}\right) \, 
\Gamma(1+\epsilon) \, (1+\beta^2)^2 \, \int_0^1 dz 
\nonumber \\ && \hspace{-10mm}
\times \int_0^1 \frac{dy \, (1-y)^{-\epsilon}}{\left[2\beta^2(1-y)^2 z^2
-2\beta^2(1-y)z-\frac{(1-\beta^2)}{2}\right]^{1-\epsilon/2}}
\int \frac{d^n k_1}{k_1^2 \, v_i \cdot k_1 \, 
\left[\left((v_i-v_j)z+v_j\right)\cdot k_1\right]^{1+\epsilon}} \, .
\nonumber 
\label{I2}
\eeqa
Now we proceed with the $k_1$ integral and separate the UV and IR poles.
After many steps, we find the $1/\epsilon^2$ and $1/\epsilon$ UV poles of $I_2$:
\beqa
I_2^{UV}&=&-\frac{\alpha_s^2}{\pi^2} \frac{(1+\beta^2)^2}{8\beta^2}
\frac{1}{\epsilon} \left\{\ln\left(\frac{1-\beta}{1+\beta}\right)
\left[2\, {\rm Li}_2\left(\frac{2\beta}{1+\beta}\right)
+4\, {\rm Li}_2\left(\frac{1-\beta}{1+\beta}\right) \right. \right.
\nonumber \\ && \hspace{55mm} \left.
{}+2\, {\rm Li}_2\left(\frac{-(1-\beta)}{1+\beta}\right) 
-\ln(1+\beta) \, \ln(1-\beta)-\zeta_2\right]
\nonumber \\ &&  \hspace{2mm} \left.
{}-2\ln^2\left(\frac{1-\beta}{1+\beta}\right)
\ln\left(\frac{1+\beta}{2\beta}\right)
+\frac{1}{3}\ln^3(1-\beta)-\frac{1}{3}\ln^3(1+\beta)
-{\rm Li}_3\left(\frac{(1-\beta)^2}{(1+\beta)^2}\right)+\zeta_3 \right\}.
\nonumber
\label{I2final}
\eeqa
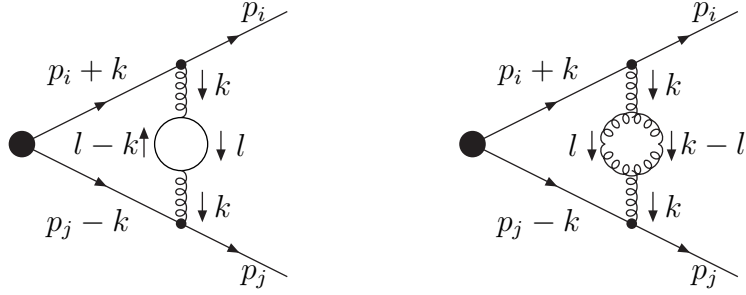
\begin{figure}[htb]
\begin{center}
\begin{picture}(120,120)(0,0)
\Vertex(0,50){5}
\ArrowLine(0,50)(60,80)
\ArrowLine(60,80)(100,100)
\Vertex(60,80){2}
\Gluon(60,80)(60,60){2}{4}
\Text(25,78)[c]{$p_i+k$}
\Text(88,100)[c]{$p_i$}
\LongArrow(68,78)(68,68)
\Text(76,73)[c]{$k$}
\LongArrow(75,55)(75,45)
\Text(83,50)[c]{$l$}
\GCirc(60,50){10}{1}
\LongArrow(46,45)(46,55)
\Text(32,50)[c]{$l-k$}
\LongArrow(68,32)(68,22)
\Text(76,27)[c]{$k$}
\ArrowLine(0,50)(60,20)
\ArrowLine(60,20)(100,0)
\Gluon(60,40)(60,20){2}{4}
\Vertex(60,20){2}
\Text(25,20)[c]{$p_j-k$}
\Text(88,0)[c]{$p_j$}
\end{picture}
\hspace{15mm}
\begin{picture}(120,120)(0,0)
\Vertex(0,50){5}
\ArrowLine(0,50)(60,80)
\ArrowLine(60,80)(100,100)
\Vertex(60,80){2}
\Gluon(60,80)(60,60){2}{4}
\Text(25,78)[c]{$p_i+k$}
\Text(88,100)[c]{$p_i$}
\LongArrow(68,78)(68,68)
\Text(76,73)[c]{$k$}
\LongArrow(75,55)(75,45)
\Text(91,50)[c]{$k-l$}
\LongArrow(45,55)(45,45)
\Text(38,50)[c]{$l$}
\GlueArc(60,50)(10,0,180){2}{6}
\GlueArc(60,50)(10,180,360){2}{6}
\LongArrow(68,32)(68,22)
\Text(76,27)[c]{$k$}
\ArrowLine(0,50)(60,20)
\ArrowLine(60,20)(100,0)
\Gluon(60,40)(60,20){2}{4}
\Vertex(60,20){2}
\Text(25,20)[c]{$p_j-k$}
\Text(88,0)[c]{$p_j$}
\end{picture}
\end{center}
\caption{Two-loop diagrams with quark and gluon loops.}
\end{figure}
We now proceed with the diagrams in Fig. 3 that involve internal quark and 
gluon loops. For the quark loop we find 
\beqa
I_{ql}=(-1) n_f g_s^4 \int\frac{d^n k}{(2\pi)^n}\frac{d^n l}{(2\pi)^n}
\frac{v_i^{\mu}}{v_i\cdot k} \frac{(-v_j^{\rho})}{(-v_j\cdot k)}
\frac{(-i)g^{\mu\nu}}{k^2} \frac{(-i)g^{\rho\sigma}}{k^2}
{\rm Tr} \left[-i \gamma^{\nu}
\frac{i l\!\!/}{l^2} (-i) \gamma^{\sigma} i \frac{(l\!\!/-k\!\!/)}
{(l-k)^2}\right] \, .
\nonumber 
\label{Iql}
\eeqa
After many steps (see also \cite{KSS}) we extract the UV poles
\beqa
I_{ql}^{UV}&=&-n_f \frac{\alpha_s^2}{\pi^2} \frac{(1+\beta^2)}{6\beta}
\left\{\frac{1}{\epsilon^2} \ln\left(\frac{1-\beta}{1+\beta}\right)
+\frac{1}{\epsilon}\left[
-{\rm Li}_2\left(\frac{1+\beta}{2}\right)
+{\rm Li}_2\left(\frac{1-\beta}{2}\right)
\right. \right.
\nonumber \\ && \hspace{10mm} \left. \left.
{}+\frac{1}{2}\ln^2(1+\beta) -\frac{1}{2}\ln^2(1-\beta)
+\left(\frac{5}{6}+4\ln 2+\ln \pi-\gamma_E-i \pi \right)
\ln\left(\frac{1-\beta}{1+\beta}\right)
\right] \right\} \, .
\nonumber 
\label{IqlUV}
\eeqa

For the gluon-loop integral, we have 
\beqa
I_{gl}&=&\frac{1}{2} g_s^4 \int\frac{d^n k}{(2\pi)^n}\frac{d^n l}{(2\pi)^n}
\frac{v_i^{\mu}}{v_i\cdot k} \frac{(-v_j^{\nu})}{(-v_j\cdot k)}
\frac{(-i)g^{\mu\mu'}}{k^2} \frac{(-i)g^{\rho\rho'}}{l^2}
\frac{(-i)g^{\sigma\sigma'}}{(k-l)^2} \frac{(-i)g^{\nu\nu'}}{k^2}
\nonumber \\ &&
\times \left[g^{\mu' \rho} (k+l)^{\sigma}+g^{\rho \sigma} (k-2l)^{\mu'}
+g^{\sigma \mu'} (-2k+l)^{\rho}\right]
\nonumber \\ &&
\times \left[g^{\rho' \nu'} (l+k)^{\sigma'}+g^{\nu' \sigma'} (-2k+l)^{\rho'}
+g^{\sigma' \rho'} (k-2l)^{\nu'}\right] \, .
\nonumber 
\label{Igl}
\eeqa
We calculate the UV poles and find
\beqa
I_{gl}^{UV}&=&-\frac{19}{96} \frac{\alpha_s^2}{\pi^2}
\frac{(1+\beta^2)}{\beta}
\left\{\frac{1}{\epsilon^2} \ln\left(\frac{1-\beta}{1+\beta}\right)
+\frac{1}{\epsilon}\left[
-{\rm Li}_2\left(\frac{1+\beta}{2}\right)
+{\rm Li}_2\left(\frac{1-\beta}{2}\right)
\right. \right.
\nonumber \\ && \hspace{10mm} \left. \left.
{}+\frac{1}{2}\ln^2(1+\beta) -\frac{1}{2}\ln^2(1-\beta)
+\left(\frac{58}{57}+4\ln 2+\ln \pi-\gamma_E-i \pi \right)
\ln\left(\frac{1-\beta}{1+\beta}\right)
\right] \right\} \, .
\label{IglUV}
\nonumber
\eeqa

We also have to add a diagram to those in Fig. 3 involving a ghost loop.
The corresponding integral is
\beqa
I_{gh}=(-1) g_s^4 \int\frac{d^n k}{(2\pi)^n}\frac{d^n l}{(2\pi)^n}
\frac{v_i^{\mu}}{v_i\cdot k} \frac{(-v_j^{\rho})}{(-v_j\cdot k)}
\frac{i}{l^2} l^{\nu} \frac{i}{(l-k)^2} (l-k)^{\sigma}
\frac{(-i)g^{\mu\nu}}{k^2} \frac{(-i)g^{\rho\sigma}}{k^2} 
\nonumber
\label{Igh}
\eeqa
and a calculation of its UV poles gives
\beqa
I_{gh}^{UV}&=& -\frac{\alpha_s^2}{\pi^2}
\frac{(1+\beta^2)}{96\beta}
\left\{\frac{1}{\epsilon^2} \ln\left(\frac{1-\beta}{1+\beta}\right)
+\frac{1}{\epsilon}\left[
-{\rm Li}_2\left(\frac{1+\beta}{2}\right)
+{\rm Li}_2\left(\frac{1-\beta}{2}\right)
\right. \right.
\nonumber \\ && \hspace{10mm} \left. \left.
{}+\frac{1}{2}\ln^2(1+\beta)-\frac{1}{2}\ln^2(1-\beta)
+\left(\frac{4}{3}+4\ln 2+\ln \pi-\gamma_E-i \pi \right) 
\ln\left(\frac{1-\beta}{1+\beta}\right)
\right] \right\}.
\nonumber
\label{IghUV}
\eeqa

We also note that the integral for another diagram 
involving an internal gluon loop 
with a four-gluon vertex vanishes.

There are additional diagrams not discussed here, 
also including self-energies and counterterms. 
The color factors for all diagrams have been calculated and must be 
accounted for in the final result.

\mysection*{Acknowledgements}

The work of N.K. was supported by the 
National Science Foundation under Grant No. PHY 0555372.

\end{document}